**Highly skewed current-phase relation in superconductor-topological insulator-superconductor Josephson junctions**

**Authors**

Morteza Kayyalha[1,*], Aleksandr Kazakov[2], Ireneusz Miotkowski[2], Sergei Khlebnikov[2], Leonid P. Rokhinson[2,1], Yong P. Chen[2,1,3,*]

**Affiliations**

[1]School of Electrical and Computer Engineering and Birck Nanotechnology Center, Purdue University, West Lafayette, IN 47907, USA

[2]Department of Physics and Astronomy, Purdue University, West Lafayette, IN 47907, USA

[3]Purdue Quantum Center, Purdue University, West Lafayette, IN 47907, USA

[*]To whom correspondence should be addressed: mkayyalh@purdue.edu, yongchen@purdue.edu

**Abstract**

Three-dimensional topological insulators (TI's) in proximity with superconductors are expected to exhibit exotic phenomena such as topological superconductivity (TSC) and Majorana bound states (MBS), which may have applications in topological quantum computation. In superconductor-TI-superconductor Josephson junctions, the supercurrent versus the phase difference between the superconductors, referred to as the current-phase relation (CPR), reveals important information including the nature of the superconducting transport. Here, we study the induced superconductivity in gate-tunable Josephson junctions (JJs) made from topological insulator $BiSbTeSe_2$ with superconducting Nb electrodes. We observe highly skewed (non-sinusoidal) CPR in these junctions. The critical current, or the magnitude of the CPR, increases with decreasing temperature down to the lowest accessible temperature ($T \sim 20$ mK), revealing the existence of low-energy modes in our junctions. The gate dependence shows that close to the Dirac point the CPR becomes less skewed, indicating the transport is more diffusive, most likely due to presence of electron/hole puddles and charge inhomogeneity. Our experiments provide strong evidence that superconductivity is induced in the topological surface states (TSS) and probes the nature of the ballistic superconducting transport in our gate-tunable TI-based JJs. Furthermore, the measured CPR is in good agreement with the prediction of a model which calculates the eigenstate energies vs. the phase in our system, taking into account TSS wavefunctions extending over the entire circumference of the TI, as well as the finite width of the electrodes.

**Main text**

**Introduction**

Three-dimensional (3D) topological insulators are a new class of quantum matters and are characterized by an insulating bulk and conducting topological surface states (TSS). These TSS are spin-helical states with linear Dirac fermion-like energy-momentum dispersion [1,2]. The TSS of 3D topological insulators (TI's) in proximity to s-wave superconductors are among the top candidates proposed to realize topological superconductors [3], capable to support Majorana bound states and promising for future applications in topological quantum computing [4,5].

A Josephson junction (JJ) made of a TI with two superconducting contacts is one of the most common platforms to study the nature of the induced superconductivity in TIs and possible topological superconductivity. One of the fundamental properties of a JJ is its supercurrent ($I$) as a function of the phase ($\varphi$) difference between two superconductors, referred to as the current-phase relation (CPR), where the maximum of $I(\varphi)$ is the critical current ($I_C$) of the JJ. Given the topological protection or the prohibited backscattering from non-magnetic impurities in the TSS of 3D TIs [1,2], superconductor-TI-superconductor (S-TI-S) junctions are expected to demonstrate novel features in their CPR. While for conventional junctions the CPR is $2\pi$-periodic, for TI-based JJs the CPR is predicted to have an additional $4\pi$-periodic component [3,6]. This $4\pi$-periodicity originates from the zero-energy crossing (at $\varphi = \pi$) of the Andreev bound states (ABS) and is protected by the fermion parity conservation. However, if the temporal variation of $\varphi$ is slower than the quasiparticle poisoning time, the $2\pi$-periodicity of the CPR is restored, which masks the unique topological nature of the Josephson junctions [6–8]. Nonetheless, in this case the topologically protected modes can give rise to highly non-sinusoidal $2\pi$-periodic CPR similar to a perfectly ballistic (scattering free) JJ [7,9].

The Josephson junctions were experimentally studied in 3D TIs [10–21] including TI nanoribbons (TINR's) [22,23]. However, it has been challenging for previous direct measurements of the CPR to show significant skewness [21,23]. In this work, we fabricate S-TI-S junctions based on the topological insulator $BiSbTeSe_2$ flakes, which have an insulating bulk and demonstrate TSS-dominated electrical properties at low temperatures [22,24,25]. We measure the CPR in the S-TI-S junctions using an asymmetric superconducting quantum interference device (SQUID) [26].

Remarkably, the measured CPR in our S-TI-S junctions are highly skewed, revealing that the superconducting transport is carried by the ballistic TSS in our TI JJs. Furthermore, we observe that the skewness depends on the back-gate voltage ($V_g$) and is the smallest close to the charge neutrality point (CNP). We present a theoretical model based on the induced superconductivity in the ballistic TSS of the TI. This model takes into account the finite-size (of both Nb and TI) and proximity effects and relates the induced supercurrent to the TSS that extend over the entire circumference of the TI. The calculated energy spectrum (energy vs. phase $\varphi$) of the junction reveals the existence of extremely low-energy modes that exist over the entire range of phases, i.e. $0 \leq \varphi < 2\pi$. The computed CPR from the theory is in excellent agreement with the experimental results.

**Materials and devices**

High quality single crystals of BiSbTeSe$_2$ were grown using the Bridgman technique as described elsewhere [24]. Such BiSbTeSe$_2$ are among the most bulk-insulating 3D TIs, where the Fermi energy lies within the bulk bandgap and inside the topological surface states (TSS), as verified by the angle resolved photoemission spectroscopy (ARPES) and transport measurements [24]. Exfoliated thin films of this material exhibit ambipolar field effect as well as several signatures of topological transport through the spin-helical Dirac fermion TSS including half-integer quantum Hall effect and $\pi$ Berry phase [24,25]. Furthermore, we have recently observed an anomalous enhancement of the critical current in BiSbTeSe$_2$ nanoribbons-based Josephson junctions, demonstrating the induced superconductivity in the TSS [23]. We obtain BiSbTeSe$_2$ flakes using the standard scotch-tape exfoliation technique and transfer them onto a 300-nm-thick SiO$_2$/500-$\mu$m-thick highly doped Si substrates, which are used as back gates. We then locate the BiSbTeSe$_2$ flakes with different width ($W$) and thickness ($t$) using an optical microscope. Subsequently, an electron beam lithography is performed to define a SQUID consisting of a TI-based JJ and a reference (REF) junction. The electrode separation, $L$, in the TI-based JJ is $\sim$ 100 nm. Finally, a thin layer ($t \sim 40$ nm) of superconducting Nb is deposited in a DC sputtering system. Prior to the Nb deposition, a brief ($\sim$ 3 seconds) in situ Ar ion milling is used to clean the interface between Nb and the TI flake. Fig. 1a shows a scanning electron microscope (SEM) image of an asymmetric SQUID with a BiSbTeSe$_2$ flake (sample A). The data presented here comes from two devices, sample A ($W \sim 2$ μm and $t \sim 40$ nm) and sample B ($W \sim 4$ μm and $t \sim 13$ nm). All our measurements are performed in a dilution refrigerator with a base temperature ($T$) of $\sim 20$ mK.

**CPR measurement and discussion**

We adapt an asymmetric SQUID technique [26] to measure the CPR in our TI (BiSbTeSe$_2$) based JJ. The asymmetric SQUID consists of two Josephson junctions in parallel as shown by the SEM image in Fig. 1a. The first JJ is the S-TI-S junction with an unknown CPR, $I(\varphi)$, and is highlighted by the dashed white rectangle in Fig. 1a. The second JJ is a conventional S-S'-S junction (REF junction), where S and S' are 300-nm and 80-nm wide Nb lines, respectively. Supercurrent ($I^{REF}$) in the REF junction follows a sinusoidal behavior vs. the phase difference ($\varphi_R$) across the junction, hence $I^{REF}(\varphi_R) = I_C^{REF} \sin(\varphi_R)$, where $I_C^{REF}$ is the critical current of the REF junction. The total current ($I^{SQUID}$) of the SQUID device is $I^{SQUID} = I^{REF}(\varphi_R) + I(\varphi)$. Furthermore, the phase differences across the two JJs and the external magnetic flux $\Phi_B = B \cdot S$, where $S$ is the area of the SQUID, are related by $\varphi - \varphi_R = 2\pi \frac{\Phi_B - L_S I}{\Phi_0}$, where $\Phi_0 = \frac{h}{2e}$ is the superconducting magnetic flux quantum, and $L_s$ is the self-inductance of the SQUID loop. We can estimate $L_S = \hbar R_N / \pi \Delta_0$ = 70 pH [27], where $R_N \sim 100$ Ω is the normal-state resistance of the SQUID (measured slightly above its critical temperature $T_C \sim 2$ K), $\Delta_0 = 1.76 k_B T_C \sim 0.3$ meV is the zero-temperature superconducting gap (calculated from the BCS theory) of the junction, and $k_B$ is the Boltzmann constant. Since $L_S I / \Phi_0 \sim 0.07 \ll 1$, we can ignore the contribution of $L_S$ in the phase difference, thus $\varphi - \varphi_R = 2\pi \frac{\Phi_B}{\Phi_0}$ and $I^{SQUID} = I^{REF}(\varphi_R) + I\left(2\pi \frac{\Phi_B}{\Phi_0} + \varphi_R\right)$. Furthermore, the REF junction is designed such that $I_C^{REF} \gg I_C$, the critical current in the S-TI-S junction, thus $\varphi_R = \frac{\pi}{2}$ and $I_C^{SQUID} \sim I_C^{REF} + I\left(2\pi \frac{\Phi_B}{\Phi_0} + \frac{\pi}{2}\right)$. Therefore, the modulation (in period of $B_0 = \Phi_0 / S$) of the $I_C^{SQUID}$ vs. $B = \Phi_B / S$ will directly probe $I(\varphi)$ or the CPR of the TI-based JJ.

Fig. 1b depicts the set-up for the measurement of the CPR in our TI-based JJs. In order to reduce the uncertainty of the measured $I_C^{SQUID}$ due to thermal and quantum fluctuations, we use a square wave pulsed current (frequency $f \sim 17$ Hz) with 50% duty cycle to bias the SQUID. The voltage ($V_S$) across the SQUID is measured by a lock-in amplifier and the magnetic flux in the SQUID is varied by an external magnet. For a fixed $\Phi_B$, once the amplitude of the pulsed current is increased above $I_C^{SQUID}$, a non-zero voltage appears across the SQUID. Fig. 1c depicts a color map of $V_S$ as functions of $I^{SQUID}$ and the external magnetic field ($B$) applied out of the plane of the SQUID. The solid white curve highlights $I_C^{SQUID}$ vs. $B$. We estimate $I_C^{REF} \sim 18$ μA (the dashed red line in Fig. 1c) by taking average of $I_C^{SQUID}$ vs. $B$. Then $I(\varphi) \sim I_C^{SQUID}(2\pi \cdot \Phi_B/\Phi_0) - I_C^{REF}$. Fig. 1d. shows the supercurrent $I(\varphi)$ normalized by its amplitude ($I_C$) vs. $\varphi$ measured in sample A for $V_g = 0$ V at $T = 20$ mK (red symbols). Since the absolute value of the flux in the SQUID is unknown (for instance, due to a remnant field), we shift the experimental curve in the horizontal direction such that $\varphi = 2\pi \cdot \Phi_B/\Phi_0$. The measured CPR is contrasted with a reference $\sin(\varphi)$ shown by the dashed blue curve in Fig.1d. The measured CPR in sample A is forward skewed, i.e. its maximum occurs at $\varphi = 0.75\pi$ (instead of $\pi/2$ for $\sin(\varphi)$).

Fig. 2a depicts the normalized supercurrent $I/I_C$ vs. $\varphi$ measured at a few different temperatures in sample A. The amplitude of the CPR (i.e. $I_C$) as a function of $T$ is plotted in Fig. 2b. We observe that the CPR remains highly non-sinusoidal up to $T \sim 400$ mK, but becomes nearly sinusoidal at higher $T = 1.3$ K. Furthermore, $I_C$ exhibits a strong $T$ dependence and increases as we decrease the temperature down to the lowest accessible $T = 20$ mK. Such a temperature dependence is in contrast to that of conventional junctions, where $I_C$ is expected to saturate at low temperatures [28]. Fig. 2c depicts the amplitude of the fast Fourier transform (FFT) normalized by the amplitude of the first harmonic as a function of $\frac{2\pi}{\varphi} = \frac{B_0}{B}$, where $B_0 = \Phi_0/S \sim 1.1$ Gauss and $S \sim 16$ μm$^2$ is the area of the SQUID. The FFT is calculated for the data taken at $T = 20$ mK in the range $-5 \le \varphi/2\pi \le 5$ and reveals that the CPR can be described by a Fourier series with up to six harmonics. The blue and black curves are predictions of a general model for ballistic junction and our model for TI junction, respectively, and will be discussed later. In order to describe the shape of the CPR in our samples, we define the total harmonic distortion ($THD$) as

$$THD = \sqrt{\frac{\sum_{j=2}^6 A_j^2}{A_1^2}},\tag{1}$$

where $A_j$ is the amplitude of the $j^{th}$ harmonic. Fig. 2d depicts $THD$, $A_2/A_1$, and $A_3/A_1$ vs. $T$ in sample A at $V_g = 0$ V. We observe that $THD$, $A_2/A_1$, and $A_3/A_1$ are nearly temperature independent up to $T \sim 400$ mK. Moreover, at $T = 1.3$ K, $A_3/A_1 \sim 0$ and $THD \sim A_2/A_1$, indicating that at this temperature, only the first and second harmonics are present in the CPR. Thus, the CPR of the TI junction is less skewed compared to that at the base temperature.

Fig. 3a demonstrates the CPR measured at different $V_g$'s for sample B at $T = 30$ mK. The inset of Fig. 3b depicts the two-terminal resistance $R$ of the SQUID vs. $V_g$ measured at $T = 8$ K, above the critical temperature ($T_c^{Nb} \sim 7$ K) of the Nb electrodes. The charge neutrality point (CNP) in this sample is at $V_{CNP} \sim$ -15 V. Compared to sample A, sample B exhibits a stronger gate dependence and an ambipolar field-effect in its normal-state resistance. We also observe that in sample B the skewness changes as a function of $V_g$. Fig. 3b plots the $THD$ vs. $V_g$ for both sample A (red) and sample B (blue). We note that the CPR is most skewed in sample B at $V_g = 30$ V, where the chemical potential is inside the bulk bandgap yet away from the CNP (see the inset of Fig. 3b). The reduced skewness at $V_g \sim 0$ V may be a result of the charge inhomogeneity and electron/hole puddles near the CNP.

**Theoretical Model**

In this section, we introduce a theoretical model based on the induced superconductivity in the spin-helical surface states of topological insulators. Since the superconducting (Nb) contacts in our case are only 300 nm wide (a value comparable to the expected coherence length $\xi = \hbar v_F / \Delta_0$ ~330 nm of the junction), we cannot assume existence of the Andreev bound states (confined inside the junction) but should suppose instead that the surface state wavefunction extends over the entire circumference of the sample (see Fig. 4a). We denote the circumference $C_x$ and define the longitudinal coordinate $x$ to be in the range $-\frac{C_x}{2} \leq x \leq \frac{C_x}{2}$. We adopt the Hamiltonian of Fu and Kane [3] and take the pairing amplitude to be a piecewise constant function of $x$ as follows: $\Delta(x) = \Delta_0 e^{\frac{i\varphi}{2}}$ for $\frac{L}{2} < x < \frac{L}{2} + b$, $\Delta_0 e^{\frac{-i\varphi}{2}}$ for $-\frac{L}{2} - b < x < -\frac{L}{2}$, and zero otherwise. Here $L$ and $b$ are the separation and width of the contacts, respectively. The wavefunction is subject to antiperiodic boundary conditions in $x$ [29]. In this simple model, we assume that the system is translationally invariant in the $y$ direction, so the wavefunction depends on $y$ as $\exp(ik_y y)$ for some $k_y$. This renders the problem effectively one-dimensional.

Our explanation of the CPR and temperature dependence of the critical current is based on an interplay between the finite-size and proximity effects. To compute the CPR, we first rewrite, following Ref. [3], the Hamiltonian of the surface fermions as $H = \left(\frac{1}{2}\right) \Psi^{\dagger} \{\mathcal{H}\} \Psi$, where $\Psi$ is an extended (four component) fermion multiplet and $\{\mathcal{H}\}$ is a 4 by 4 matrix in the component space. For given values of $k_y$ and chemical potential $\mu$, the $x$ component of the wavenumber, $k_x$, at the Fermi surface is $k_x = \pm k'$, where $\hbar v_F k' = \sqrt{\mu^2 - \hbar^2 v_F^2 k_y^2}$, and $v_F$ is the Fermi velocity of the TSS. We choose two momentum ranges, each consisting of $j = 60\text{-}200$ momenta, one range around $k'$ and the other symmetrically to it around $-k'$ and consider the Fourier expansion of $\Psi$ in the corresponding set of plane waves. Components of $\Psi$ with different values of $k_x$ are connected by the Fourier transform of the pairing amplitude $\Delta(x)$. This converts the eigenvalue problem for $\{\mathcal{H}\}$ into a matrix problem, which we diagonalize numerically for various values of the phase difference $\varphi$. $\{\mathcal{H}\}$ has a particle-hole symmetry, which stems from using four fermionic components in place of two: at each $\varphi$, the energy levels come in $\pm E$ pairs. In terms of the nonnegative levels $E_n \geq 0$ (one from each pair), the total free energy at finite temperature $T$ is:

$$F(\varphi) = -\frac{1}{2} \sum_n E_n(\varphi) - k_B T \sum_n \ln\left[1 + e^{-\frac{E_n(\varphi)}{k_B T}}\right], \qquad (2)$$

and the current is obtained as $I(\varphi) = \left(\frac{2e}{\hbar}\right) \frac{dF}{d\varphi}$. As we increase the number of $k_x$ (or $j$) participating in the expansion, $F(\varphi)$ suffers from an ultraviolet divergence, but the current does not. To calculate finite temperature properties, we replace $\Delta_0$ above with the $T$-dependent superconducting gap $\Delta(T)$ modeled using the BCS self-consistent equation [30].

The energy spectrum ($\pm E_n$ vs. $\varphi$) for sample A for the modes within the gap, $|E_n| \leq \Delta_0$ is shown in Fig. 4b. Interestingly, we observe modes with energies much smaller than $\Delta_0$ that extend over the entire range of $\varphi$, see red curves in Fig. 4b. These low-energy states lead to the non-saturation of the junction's critical current down to our lowest accessible temperature ($T \sim 20$ mK) as seen in Fig. 2b in the theoretical (blue) curve, consistent with the experimental data (symbols).

Because the wavefunction extends over the entire circumference $C_x$, while the Nb contacts occupy only a small part of it, the energy scale of the low-energy modes is only a fraction of the full $\Delta_0$. Our results can be understood qualitatively using the perturbation theory. For $\Delta_0 = 0$, energy $E =$ $\pm \left| \hbar v_F \sqrt{k_x^2 + k_y^2} \pm \mu \right|$ [3], so there is a strictly zero energy state whenever $k'$, defined above, equals one of the quantized free-fermion momenta

$$k' = \frac{2\pi}{C_x}\left(n + \frac{1}{2}\right),$$   (3)

where $n \geq 0$ is an integer. When $\Delta_0 > 0$, these states are gapped roughly by $2b\Delta_0/C_x$. For sample A with $C_x \sim 6$ μm and the contact width $b \sim 300$ nm, this is about $0.1\ \Delta_0$. Crucially, these low-energy states exist for the entire range of phases, $0 \leq \varphi < 2\pi$, in contrast for instance to the case of a conventional ballistic junction (the Kulik-Omelyanchuk theory [31]), where the minimal excitation energy remains on the order $\Delta_0$ except for a narrow vicinity of $\varphi = \pi$.

For a given $\mu$, the condition (3) will be satisfied better for some $k_y$ than for others. In practice, the translational invariance in the $y$ direction is not precise, so $k_y$ is not an exact quantum number. Nevertheless, because of the large size of the TI flake in the transverse ($y$) direction, the quantization interval for $k_y$ is small, so unless $\mu$ is exceptionally close to zero, we expect there will be a significant number of modes for which (3) is satisfied to a good accuracy. We therefore adopt the simple model in which we calculate the supercurrent for $k_y = 0$ only and multiply the result by an effective number of transverse channels $N_{ch}$ to account for the contribution of all the modes. We determine $N_{ch}$ by matching the overall magnitudes of the experimental and computed critical currents. We find $N_{ch} \sim 19$ and $N_{ch} \sim 46$ for sample A at $V_g = 0$ V and sample B at $V_g = 30$ V, respectively.

We plot the computed CPR for sample A as solid curves in Fig. 2a, where an excellent agreement with the measured data is observed. The blue curve in Fig. 2c is the FFT calculated for the theoretical CPR (in the range $-5 < \varphi/2\pi < 5$ and at $T = 20$ mK) of a perfectly ballistic short junction [28,32]:

$$I(\phi) = \Delta(T) \sin\left(\frac{\varphi}{2}\right) \tanh\left(\frac{\Delta(T)\cos\left(\frac{\varphi}{2}\right)}{2k_B T}\right),$$   (4)

where $\Delta(T)$ is the temperature-dependent superconducting gap of the junction obtained from the BCS theory [30]. Notably, the experimentally observed $A_2/A_1$ in Fig. 2c is within 3% of that predicted for the fully ballistic junction (blue curve) and the $THD = 0.46$ extracted from our measured CPR at $T = 20$ mK is within 20% of the theoretical ballistic limit ($THD = 0.55$), indicating superconducting transport is nearly ballistic in sample A. The black curve in Fig. 2c plots the FFT of the CPR calculated using our theoretical model (Fig. 4). We observe that the FFT of the CPR, calculated using our model, is in reasonable agreement with the FFT of the measured CPR. In contrast, the perfectly ballistic model (blue curve) notably overestimates the higher harmonics ($A_3$ and above). The computed CPR for sample B is plotted with dashed curves in Fig. 3a at two different $V_g$'s corresponding to $\mu = 0$ and 50 meV, respectively. While the theoretical CPR agrees well with the experiment for $V_g = 30$ V, we see a deviation between the theory and experiment for $V_g = 0$ V. Sample B is much thinner ($t \sim 13$ nm) than sample A ($t \sim 40$ nm). When a TI becomes sufficiently thin, there may be a gap opening in the TSS close to the Dirac point due to hybridization of the top and bottom surface states. This gap causes the TI to transition into a trivial insulator. Moreover, there are electron-hole puddles and charge inhomogeneity near the charge neutrality point. Therefore, the transport may be more diffusive, i.e. the CPR is more sinusoidal, close to the CNP due to effects of disorder and hybridization. Such effects are not included in our theory and may be responsible for the discrepancies between the calculated and measured CPR at $V_g = 0$ V in sample B.

We note that in our previous experiments on S-TINR-S JJs [23], even though we have also observed evidence that the superconductivity is induced in the TSS, we only observe a sinusoidal CPR. A possible reason for this is a much smaller transverse size ($C_y$) of the TINR compared to the flakes used in the current work. As a consequence, $k_y$ is quantized in larger units (i.e. $2\pi/C_y$), and the condition (3) is less readily satisfied. Effectively, the small transverse size generates a gap in the TSS spectrum, preventing occurrence of low-energy states and rendering the CPR more sinusoidal at our experimental temperatures [23]. A similar explanation may be relevant also for sample B of the present paper near the charge neutrality point.

## Conclusions

We have measured the CPR, one of the fundamental properties of a Josephson junction, in a BiSbTeSe$_2$-based JJ using an asymmetric SQUID technique. We observed highly forward-skewed CPR, indicating that the superconducting transport through the TSS of the TI junction was close to ballistic. Temperature and gate dependence of the CPR were also studied, where we observed that CPR became more sinusoidal at high temperatures ($T \sim 1.3$ K) and close to the CNP. The reduced skewness near CNP was an indication of diffusive transport and was associated with the existence of electron-hole puddles and charge inhomogeneity in the very thin TI. Moreover, we developed a theoretical model that considered induced superconductivity in the spin-helical TSS of TIs. Our model assumed that the surface states can extend over the entire circumference of the TI. The predicted skewness of the CPR and the dependence on the temperature were consistent with our experimental observations. Overall, the experiment and the theory pointed toward robust features that made our TI system an excellent candidate to observe topological superconductivity and Majorana bound states.

**Acknowledgement**

M.K., A.K., L.P.R., and Y.P.C. acknowledge support from National Science Foundation (NSF) under Award DMR-1410942. M.K. and Y.P.C. acknowledge partial support from NSF under Award EFMA-1641101. L.P.R. and A.K. also acknowledge partial support from the U.S. Department of Energy (DOE), Office of Basic Energy Sciences (BES) under Award DE-SC0008630.

**Figure Captions**

Fig. 1. **Measurement of CPR using asymmetric SQUID.** (a) A scanning electron microscope (SEM) image of an asymmetric superconducting quantum interference device (SQUID) used to measure the current-phase relation (CPR), the supercurrent $I$ vs. $\varphi$ (the phase difference between the superconductors), in the topological insulator (TI)-based Josephson junction (JJ). The asymmetric SQUID is formed between a TI-based JJ with superconducting Nb contacts and a reference (REF) junction in parallel. The REF junction is a conventional S-S'-S Josephson junction with the supercurrent $I^{REF}(\varphi_R) = I_C^{REF}\sin(\varphi_R)$, where $I_C^{REF}$ and $\varphi_R$ are the critical current and the phase difference across the REF junction, respectively. (b) Schematic of the CPR measurement setup. We use a low-frequency (~17 Hz) square-wave pulsed current ($I^{SQUID}$) with 50% duty cycle to bias the SQUID. The voltage $V_S$ across the SQUID is monitored with a lock-in amplifier. A perpendicular magnetic field $B$ is applied to control the phase difference inside the SQUID loop (with area $S$). i.e. $\varphi - \varphi_R = 2\pi\,\Phi_B/\Phi_0$, where $\Phi_B = B \cdot S$ is the magnetic flux and $\Phi_0 = h/2e$ is the superconducting flux quantum. (c) Color map of $V_S$ as functions of $I^{SQUID}$ and $B$. The solid white curve marks the critical current $I_C^{SQUID}$ of the SQUID and the dashed red line is the critical

current $I_C^{REF}$ of the REF junction (d) The current-phase relation (symbols) represented by the normalized current ($I/I_C$) of the TI-based JJ vs. the phase $\varphi$ measured in sample A at temperature $T = 20$ mK. Dashed blue curve depicts $\sin(\varphi)$. Since the absolute value of the flux inside the SQUID is unknown, we shift the experimental curve in the horizontal axis so that $\varphi = 2\pi \cdot \Phi_B/\Phi_0$.

Fig. 2. **Temperature-dependence of CPR.** (a) The CPR measured for different temperatures. Symbols are experimental data and solid curves are theoretical calculations. All curves are shifted vertically for clarity. (b) Temperature dependence of the critical current ($I_C$, the amplitude of the CPR). Solid blue curve is the theoretical calculation. (c) The amplitude of the fast Fourier transform (FFT, red curve) of the CPR measured over $-5 < \varphi/2\pi < 5$ at 20mK, normalized to the amplitude ($A_1$) of the first harmonic vs. $2\pi/\varphi = B_0/B$, where $B_0 = \Phi_0/S \sim 1.1$ Gauss and $S = 16 \text{ μm}^2$. Black and blue curves are FFT's of the calculated CPR using our theoretical model and the perfectly ballistic model (Eq. 4), respectively. (d) Total harmonic distortion ($THD$) and the normalized amplitude of the second ($A_2/A_1$) and third ($A_3/A_1$) harmonics vs. $T$, where $A_j$ is the amplitude of the $j^{th}$ harmonic. All data in this figure are measured in Sample A at the gate voltage $V_g = 0$ V. Theoretical calculations are performed for $L = 100$ nm, $b = 300$ nm, $C_x = 6$ μm, $\Delta_0 = 0.3$ meV, $\mu = 50$ meV, and $\hbar v_F = 1$ eVÅ.

Fig. 3. **Gate-dependence of CPR.** (a) The CPR measured in sample B at $T = 20$ mK for different $V_g$'s. Curves are shifted vertically for clarity. Dashed black and red curves are theoretically calculated CPR with chemical potential $\mu = 0$ and 50 meV, respectively. (b) The total harmonic distortion $THD$ of the CPR as a function of $V_g$ for samples A (red) and B (blue) at $T = 20$ mK. Inset: two-terminal resistance $R$ of the SQUID (containing the parallel contribution of the TI JJ resistance and the REF JJ resistance) vs. $V_g$ at $T \sim 8$ K above the critical temperature of Nb electrodes ($T > T_c^{Nb} \sim 7$ K).

Fig. 4. **Theoretical Model.** (a) Schematic representation of the TI-based Josephson junction in our theoretical model. We assume the TSS wavefunctions extend over the entire circumference of the TI as shown by the black circle in this figure. The pairing amplitude $\Delta$ is a piecewise constant function of $x$, as follows: $\Delta(x) = \Delta_0 e^{\frac{i\phi}{2}}$ for $\frac{L}{2} < x < b + \frac{L}{2}$, $\Delta_0 e^{\frac{-i\phi}{2}}$ for $-\frac{L}{2} - b < x < -\frac{L}{2}$, and zero

otherwise. (b) Energy spectrum (energy $E$ vs. phase $\varphi$) of the modes (the lowest energy ones are highlighted with red) within the superconducting gap $\Delta_0$ computed using Eq. (2) and parameters ($L = 100$ nm, $b = 300$ nm, $C_x = 6$ μm, $\Delta_0 = 0.3$ meV, $\mu = 50$ meV, and $\hbar v_F = 1$ eVÅ) of sample A.

**Figure 1**

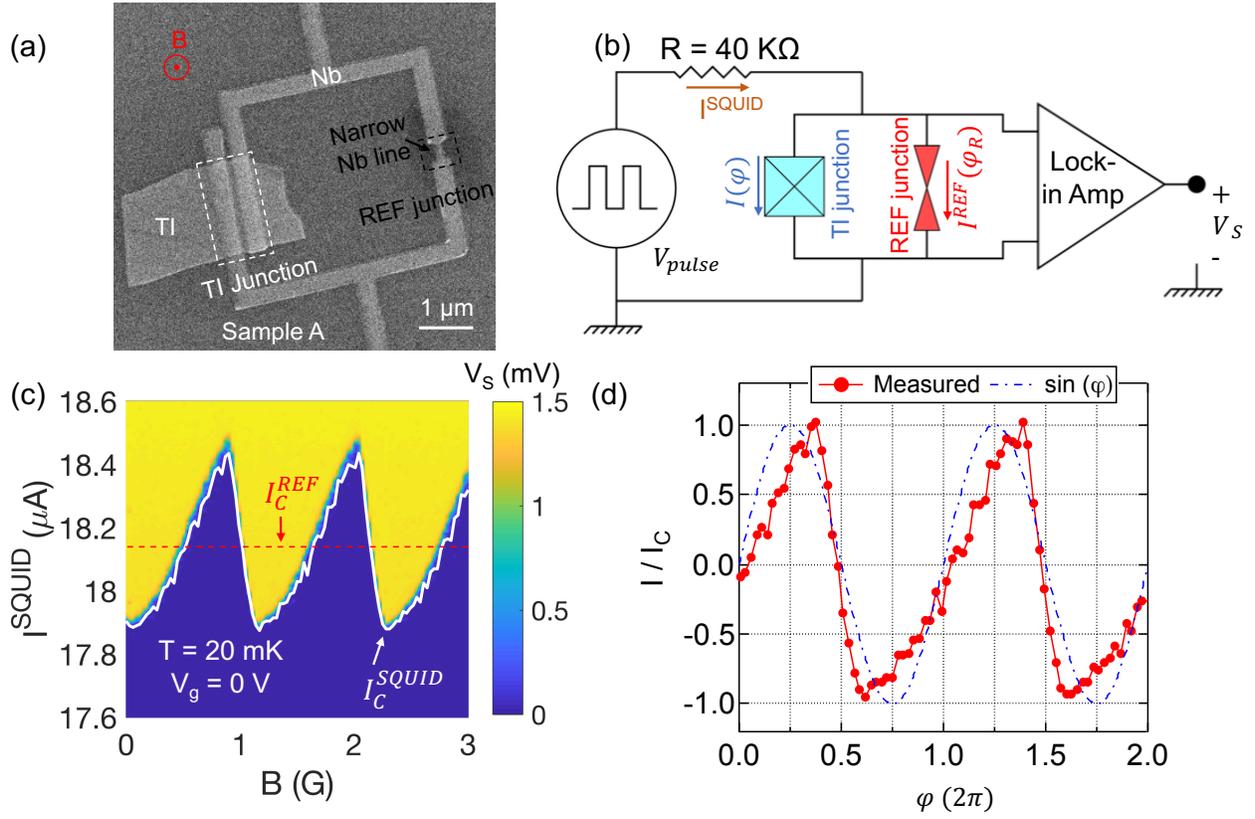

(a) [SEM image labeled: B, Nb, Narrow Nb line, REF junction, TI, TI Junction, Sample A, 1 μm]

(b) [Circuit diagram: R = 40 KΩ, $I^{SQUID}$, $V_{pulse}$, $I(\varphi)$ TI junction, REF junction, $I^{REF}(\varphi_R)$, Lock-in Amp, $V_S$]

(c) [Plot: $I^{SQUID}$ (μA) vs B (G), $V_S$ (mV) color scale 0 to 1.5, $I_C^{REF}$, $I_C^{SQUID}$, T = 20 mK, $V_g$ = 0 V]

(d) [Plot: $I / I_C$ vs $\varphi$ ($2\pi$), Measured, sin($\varphi$)]

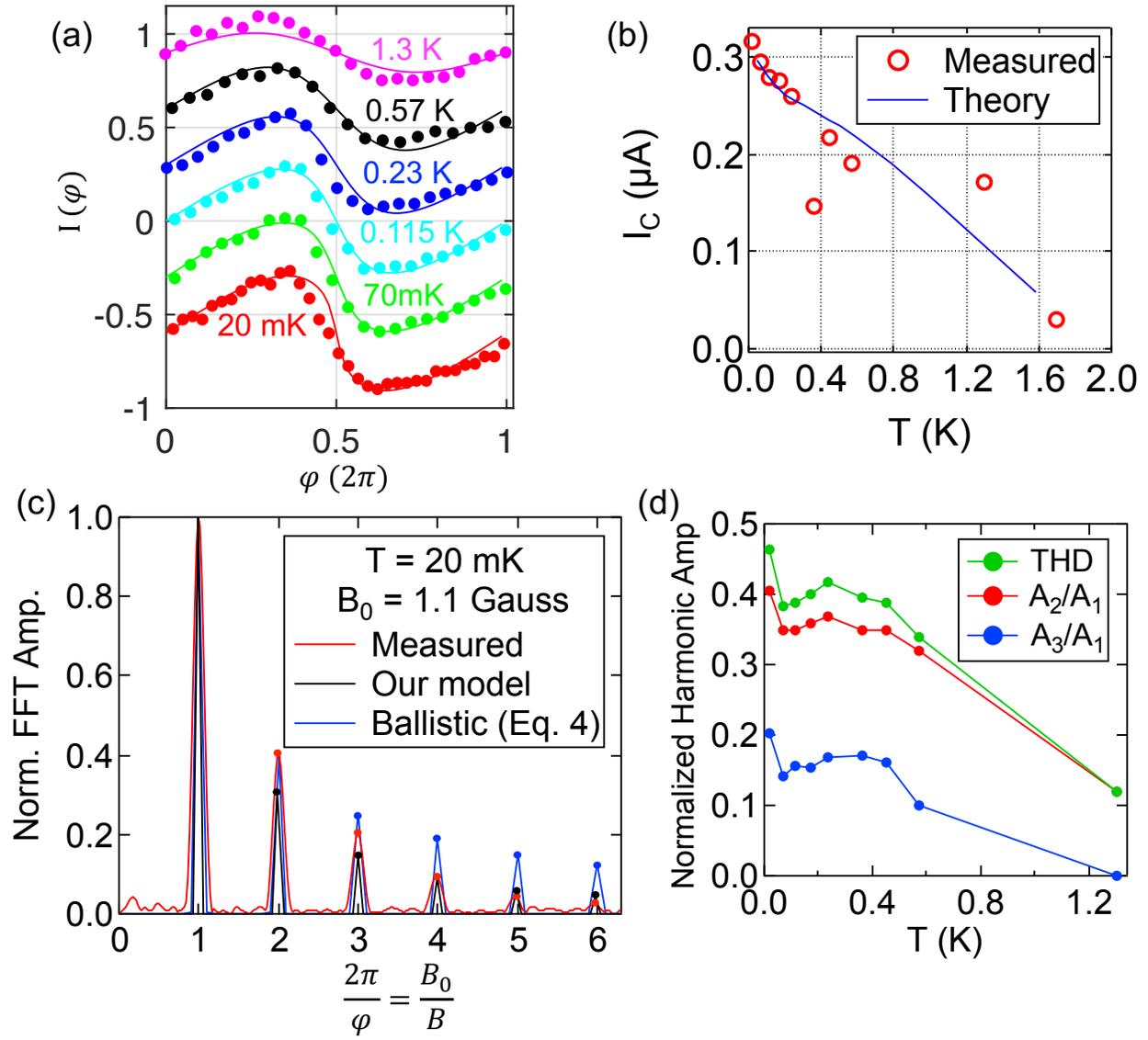

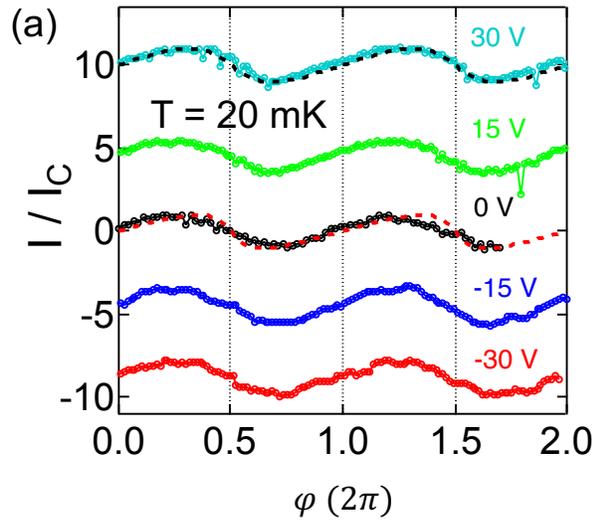

(a)

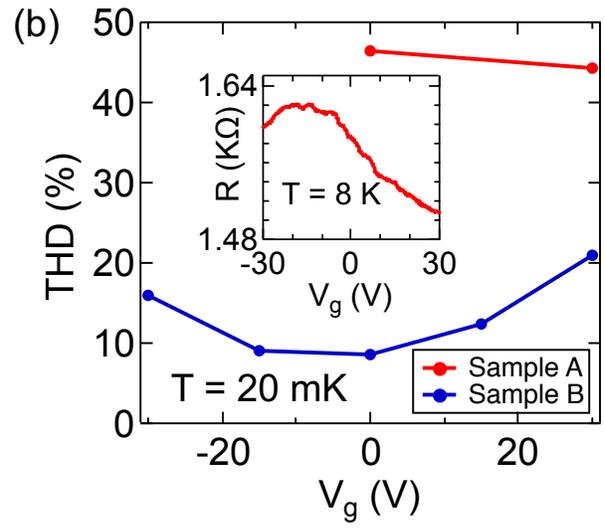

(b)

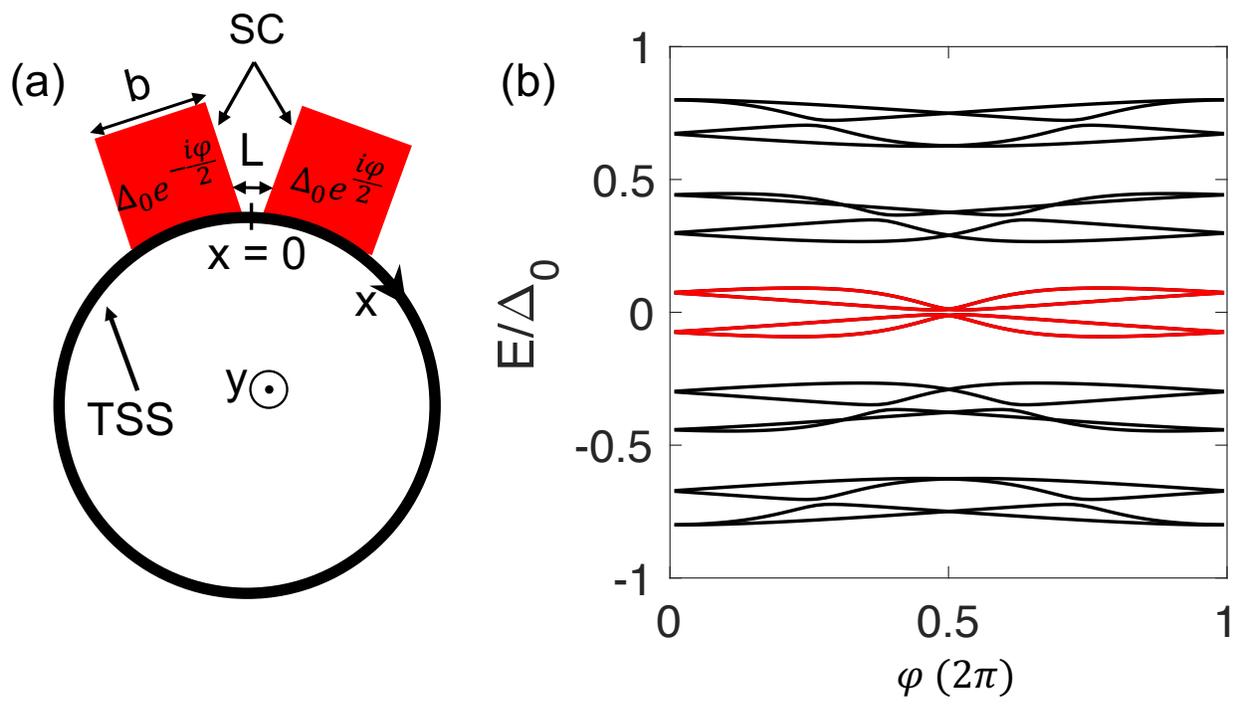